# An Agile Software Engineering Method to Design Blockchain Applications

Michele Marchesi[†]
DMI
University of Cagliari
Cagliari, Italy
marchesi@unica.it

Lodovica Marchesi
DIEE
University of Cagliari
Cagliari, Italy
lodo.marchesi@gmail.com

Roberto Tonelli
DMI
University of Cagliari
Cagliari, Italy
roberto.tonelli@dsf.unica.it

## ABSTRACT

Cryptocurrencies and their foundation technology, the Blockchain, are reshaping finance and economics, allowing a decentralized approach enabling trusted applications with no trusted counterpart. More recently, the Blockchain and the programs running on it, called Smart Contracts, are also finding more and more applications in all fields requiring trust and sound certifications. Some people have come to the point of saying that the "Blockchain revolution" can be compared to that of the Internet and the Web in their early days. As a result, all the software development revolving around the Blockchain technology is growing at a staggering rate. The feeling of many software engineers about such huge interest in Blockchain technologies is that of unruled and hurried software development, a sort of competition on a first-come-first-served basis which does not assure neither software quality, nor that the basic concepts of software engineering are taken into account.

This paper tries to cope with this issue, proposing a software development process to gather the requirement, analyze, design, develop, test and deploy Blockchain applications. The process is based on several Agile practices, such as User Stories and iterative and incremental development based on them. However, it makes also use of more formal notations, such as some UML diagrams describing the design of the system, with additions to represent specific concepts found in Blockchain development. The method is described in good detail, and an example is given to show how it works.

## 1 Introduction

Nowadays, one of the main trends of IT technology are the so-called Blockchain applications. The Blockchain is a technology originally devised to run the Bitcoin cryptocurrency in a decentralized and secure way. Subsequently, developers quickly realized that the Blockchain can be used also as a decentralized computer, running Smart Contracts – programs that can be used as the basis for automated contractual enforcements. The awareness of the potentiality of the technology – the possibility to enforce contracts getting rid of intermediaries, and space and time constraints – created a huge wave of interest in Blockchain applications. Some observers are event talking that "*we should think about the Blockchain as another class of thing like the Internet [...]*" [1] and that the "*wide adoption of Blockchain technology has the potential of reshaping the current financial services technical infrastructure.*" [2].

This interest lead to an ever increasing amount of money pouring into Blockchain initiatives. The capitalization of cryptocurrencies, despite a recent shrinking of the market, is well above 200 billion US$, and the venture capital investments, both from traditional funds and from the recent Initial Coin Offers (ICO) has overcome ten billions US$ in the past 12 months. This run to invest into new initiatives, typically quickly developing applications to be the first on the market, lead to some huge disasters, typically due to poor design and poor security practices in software development. The attacks to the so-called "Exchanges", website where it is possible to trade cryptocurrencies against each others, or even against fiat currencies like US$ and Euro, are very frequent, leading to declared losses that, summed up, amount to well over one billion US$. The feeling of many software engineers about such huge interest in Blockchain technologies and, in particular, on the many software projects rapidly born and quickly developed around the various Blockchain implementations or applications, is that of unruled and hurried software development. The scenario is that of a sort of competition on a first-come-first-served basis which does not assure neither software quality, nor that the basic concepts of software engineering are taken into account.

The first step to develop a software system using sound software engineering practices is to have a clear development process, and design practices and notations useful to the purpose [3]. Based upon this, specific development, test, deployment, and security assessment practices can be used. The goal of this paper is to propose and test a design and development process for Blockchain applications based on Smart Contracts. The overall process is mainly based on the principles of Agile Manifesto [4], complemented with some specific notation and practices. The main idea behind the proposed approach stems from the observation that a Smart Contract is a software program that runs on all the nodes of a Blockchain and whose outputs and state must be the same in all nodes. For this reason, a SC is strictly forbidden to access in anyway the external word – it can only answer to requests through a public interface, and send requests to other SCs running on the same Blockchain. Consequently, the proposed process divides the Blockchain software system specification in two parts: the specification and development of the SCs, and that of the software application(s) which interact with the external users and with the SCs. The proposed method also introduces a notation integrating the UML Use Case, Sequence, and Class diagrams, to account for Blockchain specificities.

The proposed approach has been tested on some real projects, carried on in our University and in a spinoff firm of the same University.

The remainder of this paper is organized as follows. In Section 2 we present the related work in the same, or similar fields. Section 3 describes the proposed process, and the modifications of some UML diagrams to cope with SC concepts. A simplified example, drawn from a real case, is presented in Section 4. Finally, Section 5 presents the conclusions and future work ideas.

## 2  Related Work

The research field of design methods, and in general of software engineering practices aimed at Blockchain-Oriented Software (BOS) development is still in its infancy. The first call for BOS Engineering is the paper by Porru et al. [3], who advocate the study and development of sound engineering practices to ensure effective testing activities, enhance collaboration in large teams, and facilitate the development of Smart Contracts. They also argue that that existing design notations could be adapted to better specify BOS. Xu et al. [5] present a taxonomy of Blockchain concepts based on Blockchain properties, and propose a flowchart and an initial checklist helping design decisions. Wessling et al. [6] propose an approach to decide which elements of an application architecture could benefit from the use of Blockchain technology. They identify participants, their trust relations and interactions to derive an architecture embedding Blockchain technology in existing software systems or creating new systems using Blockchain only in certain parts. Fridgen et al. [7] apply an action design research approach and situational method engineering to propose a method for the development of Blockchain use cases. They evaluated the method in four distinct industries: banking, insurance, construction and automotive.

Regarding possible extensions to the Unified Modeling Language [8] to cope with BOS, several papers have been published to propose extensions to UML notation to make it able to better represent specific fields. Baumeister et al. Proposed and extension of UML for Hypermedia design, adding new stereotypes and a new Navigational Structure Model [9]. Also Baresi et al. integrate in a single framework structural and navigational abstractions, introduced by hypermedia web models, with functional and behavioral primitives provided by UML [10]. More recently, Rocha and Ducasse [11] show three complementary modeling approaches based on well-known software engineering models – E-R diagrams, UML and BPMN – and apply them to a Smart Contract design example. Regarding UML, they propose some improvements to UML Class Diagram to better represent Smart Contract concepts.

## 3  Background

We define as Blockchain-oriented Software (BOS) all software working with an implementation of a Blockchain. A Blockchain is a distributed data structure characterized by the following key elements:

- data redundancy (each node has a copy of the Blockchain);
- check of transaction requirements before validation;
- recording of transactions in sequentially ordered blocks, whose creation is ruled by a consensus algorithm;
- transactions based on public-key cryptography;
- a transaction scripting language, associated to the transactions – the corresponding program is executed by all nodes, where the transaction is evaluated.

A Blockchain system is usually composed of a Blockchain, and of software interacting with it through transactions, typically providing the user interfaces to users, and possibly server activities to store additional information, and to execute business logic outside the Blockchain. Such a system is often called a dApp (decentralized application).

The software associated to the transactions, and running on the Blockchain, is usually called Smart Contract (SC), because the first envisaged applications for BOS are related to automated operations on the Blockchain, enforcing *contracts* among participants. Most of these contracts are used to manage cryptocurrencies, or tokens, having a true monetary value.

In the following of this paper, we will mainly refer to the Ethereum Blockchain, which is presently the most used to develop SCs [11]. Ethereum nodes are provided of the Ethereum Virtual Machine (EVM), able to execute a proper bytecode. In practice, Ethereum SCs are written a in specialized high-level language, called Solidity [12]. Fig. 1 shows the typical architecture running SCs.

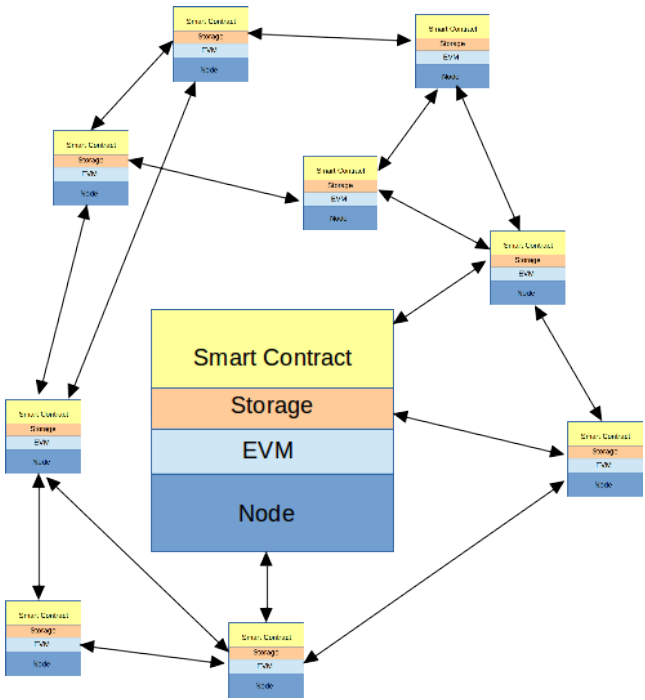

**Figure 1: Running a SC on the Ethereum Blockchain. Each node runs the same bytecode.**

A SC has a state – permanently stored in Blockchain *storage* variables. The main characteristic of the SCs is that they run in an isolated environment. The program results must be the same whatever node they run in, so, they cannot access the external world (that changes with time); they can only access and send messages to the Blockchain itself (that is immutable). On the contrary, computer programs continuously interact with the external world. Moreover, once a SC is deployed on the Blockchain, it is there forever – it cannot be undone or erased!

In Ethereum, SCs are created by special transactions; they can use other SCs, or inherit from other SCs. Creating a SC and changing its state costs GAS, which must be paid in Ether (the cryptocurrency associated to Ethereum Blockchain). A SC is endowed of public functions, that can be called after its creation (call of the constructor), or using a transaction (message call). A SC can be endowed of Ethers and can send Ethers to other SC, or to Ethereum addresses. A SC, upon a call of one of its functions, can change its state, can create and send a transaction to an address or to another SC, can call one or more of its functions, and can return a value without changing its state or sending a transactions – in this case, there is no cost for sending the message. A SC cannot initiate an action autonomously (for instance at given times), or access the external world. When developing BOS, we develop a complete system that is used by its customers, who typically do not care whether the system is based on a Blockchain or not. A BOS system is typically composed of two parts:

- a traditional software system, running on servers and/or on mobile devices, communicating with users and external devices;
- the SCs running on the Blockchain.

Our approach takes into account the substantial difference between developing traditional software and SCs, and separates the two activities. For both developments, it is based on an Agile approach. In fact, Agile methods are suited to develop systems whose requirements are not completely understood, or tend to change. These characteristics are present in dApps:

- dApps are typically very innovative applications;
- often, there is a run to write a dApp to be the first who launches it on the market.

Agile is suited for small, self-organizing teams working together, where the customer or the Product Owner (expert in the system requirements) is highly available to the team, as it is the case for many dApp teams. Moreover, Agile is iterative and incremental with short iterations, and is suited to deliver quickly and to deliver often – which is very appreciated in the context of dApp development. The key Agile practices used are requirement elicitation using User Stories (US) – short, incremental description of the functional requirements of the system from the user perspective – and iterative development implementing a subset of the USs chosen by the Product Owner at each iteration.

Other practices of Agile development that are very well suited to dApp development are: Continuous Testing, Test Driven Design, Refactoring, Continuous Integration, Collective code ownership, Information Radiators (Cards, Boards, Burndown charts), Coding Standards, Pair Programming (in some cases).

On the other hand, dApps have very strict security requirements, and a more formal approach with respect to some aspects of the development could be useful. Some key factors in SC design that must be thoroughly designed are:

- Data: permanent data are very expensive, so they must be analyzed and kept to a minimum.
- Interactions: they are key to system proper behavior, and the source of all attacks.
- Security: if there is a possible exploit, it will be exploited! Security patterns, code inspection and detailed tests must be applied to get a reasonable security level.
- Documentation: in some cases, documentation in the code is the best solution. In other cases, it is better to keep the code obscured and the documentation separated from the code.

## 3 Method

The proposed design method for BOS is performed through a sequence of steps, that are summarized in Fig. 2, in the form of a UML activity diagram.

In deeper detail, the proposed BOS development process is the following:

1. State the goal of the system, write one or two sentences summarizing the goal, and post it in a place that is visible to all developers.
2. Identify the actors which interact with the system (human roles and external systems/devices).

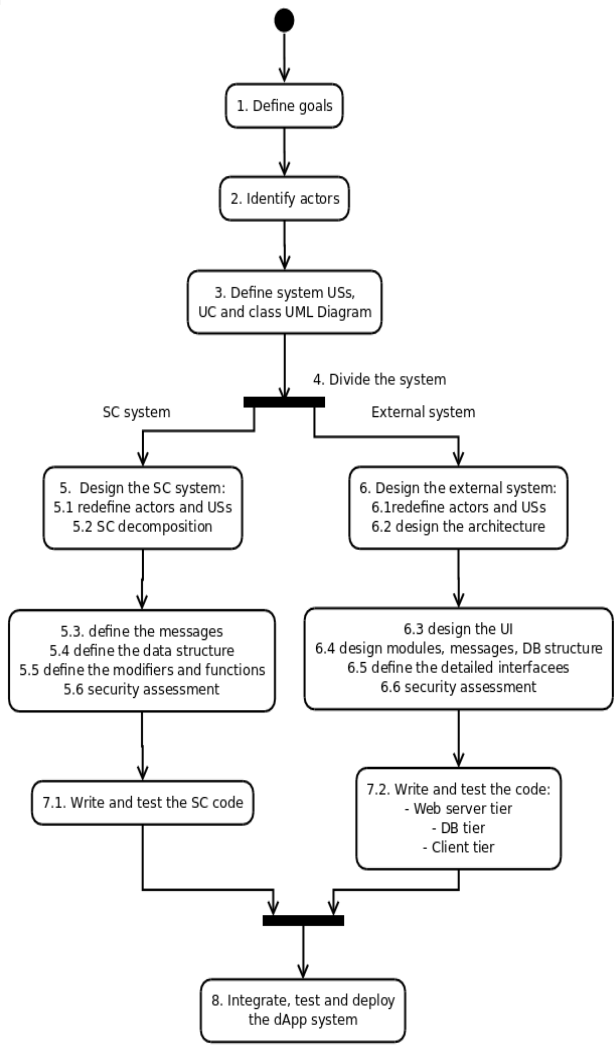

**Figure 2: The steps of the proposed BOS development process.**

Here you can possibly apply the idea of determining the trust/untrust between actors, to assess whether a Blockchain system is really needed, and for what parts [6].

3. Write the system requirements in term of user stories or features. In this phase, the system to be developed should be considered as a whole. The fact that it will be developed using a Blockchain or a set of servers in a cloud is not important.
4. Divide the system in two subsystems:
   4.1 The Blockchain system, composed by the Smart Contracts.
   4.2 The external system that interacts with the first, sending transactions to the Blockchain and receiving the results.
5. Design of the SC subsystem:
   5.1 Redefine the actors and the user stories, like those described in steps 2 and 3, considering only those directly interacting with the SC subsystem, and possible external SCs used;
   5.2 Define the decomposition in SCs (one or more); define the used libraries and external SCs; design the inheritance structure, and the usage of interfaces;
   5.3 Define the connections and the flow of messages and Ether transfers; define the state diagram (if needed);
   5.4 Define the data structure, the external interface (ABI) and the events;
   5.5 Define the internal functions and the modifiers;
   5.6 Define the tests and the security assessment practices.
6. Design of the external subsystem:
   6.1 Redefine the actors and the user stories, like those described in steps 2 and 3, but adding the new (passive) actors represented by the SC system; define the acceptance tests of the subsystem;
   6.2 Decide the broad architecture of the system, taking into account the server and client application, the Blockchain node(s) to use;
   6.3 Define the User Interface of the relevant modules, including the apps;
   6.4 Perform an analysis of the system, defining the decomposition in modules, the flow of messages, the structure and storage of permanent data, including those anchored to the Blockchain through hash digest memorization, the data or class structure of the application(s); the connections and data flows between participants, including the SCs must comply with the analysis of step 5.3;
   6.5 Define the state diagrams (if needed), the detailed interfaces of the various modules, the response to the events raised by SCs;
   6.6 Perform a security assessment of the external system.
7. Code and test the systems; in parallel:
   7.1 Write and test the SCs, starting from their data structure and functions;
   7.2 Implement the USs of external subsystem with an agile approach (Scrum or Kanban);
8. Integrate, test and deploy the overall system.

## 3.1 UML diagrams for Smart Contracts

SCs for Ethereum are typically written using the Solidity language. Solidity is an object-oriented language, and the contracts are defined in it like classes – they have a data structure, public and private functions, and can inherit from other contracts. SCs have also specific concepts like events and modifiers.

To help the modeling of SCs we use UML diagrams. However, since SCs have some very specific characteristics, we introduced some new concepts in these diagrams, to be able to better model and specify SCs. Whenever possible, these concepts are simply introduced as UML stereotypes, which are tags that can be used in UML diagrams wherever needed. In a few other cases, we had to introduce a specific notation, like the transfer of Ethers in sequence diagrams. The UML diagrams we find useful to model SCs are:

- Class diagrams, to represent the structure and relationships of SCs; we introduced various stereotypes in this kind of diagram.
- Statecharts, to represent the various states of a SC; this diagram does not need any new concept.

- Sequence diagrams, to represent the messages sent to a SC, and from a SC to another SC; this diagram needs new kind of messages – the transfer of Ethers.

UML class diagrams are used to represent SCs and structs. In Solidity, there is not the concept of *class*, but the SCs are very similar to classes. Like a class, a SC can have a data structure, public and private functions, and can inherit from one or more SCs. However, SCs have a specific nature; they are created by transactions, but a transaction can create at most a single SC. SCs, however, can send messages to other SCs, residing in the same Blockchain. In Solidity, it is also possible to define *structs*, that is complex data structures, that are not provided of functions. Consequently, the model of a SC that is created by a transaction can include other SCs it inherits from, the used structs, and the external SCs which are sent messages to.

Other, specific concepts of Ethereum SCs are *events*, flags that are raised when something relevant happens, and that signal it to the external world (which has to autonomously observe the SC, and act correspondingly), and *modifiers*, special functions that are called before a function, checking its constraints and possibly stopping the execution. Table 1 shows the stereotypes we introduced to make possible to represent SC concepts in UML class diagrams. The events could be represented in a further compartment, besides those containing the name, the attribute and the functions (operations).

**Table 1. Additions to UML class diagram (stereotypes).**

| Stereotype | Position | Description |
| --- | --- | --- |
| «contract» | Class symbol – upper compartment | Denotes a SC. |
| «library contract» | same as above | A contract taken from some (standard) library |
| «struct» | same as above | A struct, holding data but no operation, defined and used in the data structure of a contract |
| «enum» | same as above | An enum, holding just a list of possible values |
| «interface» | same as above | A contract holding only function declarations |
| «modifier» | Class symbol – lower compartment | A particular kind of function, defined in Solidity |
| «array» | Role of an association | The 1:n relationship is implemented using an array |
| «map» | same as above | The 1:n relationship is implemented using a mapping |
| «map[uint]» | same as above | The 1:n relationship is implemented using a mapping from integer to the value |

The last three stereotypes define the implementation of 1:n relationships in the data structure of a SC. In Solidity, the only supported collections to manage the storage (the data permanently stored in the Blockchain) are the array and the mapping. The former is a classical basic array of all computer languages, with the addition that new items can be added to it (but not removed). The mapping is a collection able to store key-value pairs and to efficiently retrieve a value, given its key, but not able to iterate on its elements. The last stereotype refers to a common pattern of Solidity programming – using a mapping with positive integers as its keys, so that it is possible to iterate over it.

UML Sequence Diagrams are used to model messaging. In Ethereum, the messages are associated to transactions sent to the Blockchain from external users or systems, or from SCs. Like in the object-oriented jargon, messages are synonyms of "public function calls". If a function does not write or alter the Blockchain (it is called a "view" function), the corresponding message can be sent at no cost. Other messages require to be paid in GAS to be executed.

The specificities of Ethereum regarding messaging are the kinds of participants (identified by their accounts), and the kinds of messages. The participants can be specified using stereotypes, as shown in Table 2. The messages require a specific notation.

**Table 2. Additions to UML sequence diagram (stereotypes).**

| Stereotype | Description |
| --- | --- |
| «person» | A human role, sending messages using a wallet or other application |
| «system» | An external system, able to send messages to the Blockchain |
| «device» | A device (typically IoT), able to send messages |
| «contract» | A SC, part of the system or external to it |
| «oracle» | A particular kind of SC, whose date are written by a trusted third party, and allow to access information about the external world |
| «account» | An Ethereum account, just holding Ethers. It can only receive Ethers, or send Ethers to another account or SC if the owner activates the transfer |

The different kinds of messages relevant to the design are:

- SC creation: it is sent from an external participant or from another SC; in a sequence diagram a creation is represented drawing the new participant at the time level of its creation.
- Function call: a transaction entailing Blockchain modification, and thus GAS payment; it is the classic "synchronous" or "asynchronous" message.
- View/pure function call: a transaction entailing no Blockchain modification, and no GAS payment; it can be modeled adding the «view» or «pure» stereotype to the message name.
- ETH transfers: a special transaction that transfers Ethers from an account, or a SC, to another account or SC. This is modeled using a special arrow, similar to the inheritance arrow of UML class diagrams.

## 4  An example of application

The presented SC development process is being used in our University group, and in firms we are consulting. Among the projects being developed we may quote a supply chain management system, a system to manage temporary job contracts, various remote voting systems for local authorities, and for a firm Shareholders' Meeting and board of directors meetings. Here we present a simplified version of the voting system, as an example of the first steps of the proposed process.

**Step 1. Goal of the system**. *To manage remote voting in corporate assemblies, including verification of the legal number, and proxy delegation management.*

**Step 2. Actors**. The system has basically two actors:

**Corporate administrator**: manages the system, manages the shareholders and their shares, convenes assemblies, calls for voting.

**Shareholder**: participates to assemblies, casts his votes, delegates participation to an assembly to another shareholder.

**Step 3. User Stories.** Fig. 3 shows the actors and the USs they are involved in, using a UML Use Case diagram, where the use cases are in fact USs. Note that these USs just specify the voting system, and do not depend on the specific technology used to implement it. They would be right also if the implementation did not use a Blockchain.

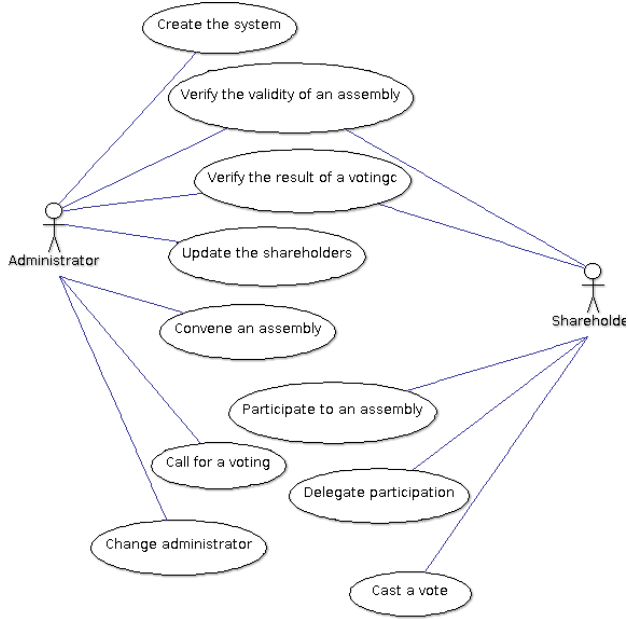

**Figure 3: The User Stories of the system specification.**

Here we have no room to show the USs in detail. Instead, in Fig. 4 we show the UML class diagram derived by an analysis of the given USs. Again, this diagram is not bound to a specific implementation of the voting system, but just shows the entities, the data structures and the operations emerging from the USs of Fig. 3.

**Step 4. Divide the system into two subsystems**. In this case the subdivision is trivial, because all USs make use of Smart Contracts.

The USs of the external app subsystem are the same. Each includes the Blockchain as a further Actor.

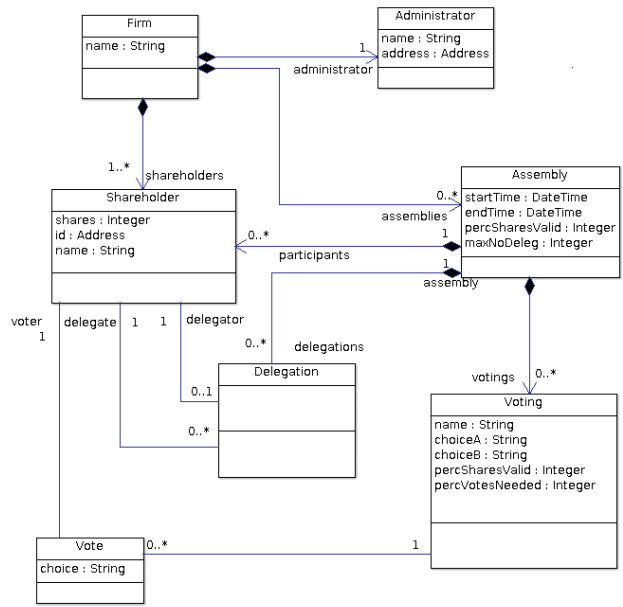

**Figure 4: The UML class diagram derived from the USs.**

The USs of the Blockchain subsystem US are the same. The identifiers of the Actors are their unique adresses:

- Corporate administrator: her/his address is at first the address that creates the contract, and then possibly a further address set by the Change administrator US.
- Shareholders: their addresses are specified and managed by the Corporate administrator.

**Step 5. Design of the SC subsystem.** The SC system is quite simple, so a single SC looks the best option. Following a well known standard, the "`Ownable`" standard abstract contract is used to manage the ownership of the Administrator on the SC:

```
contract Ownable {
   address public owner;
   modifier onlyOwner() {
      require(msg.sender == owner);
      _;
   }
}
```

The data structure of the SC is shown in Fig. 5, using a modified UML class diagram, as described in section 3.1. Fig. 6 shows the UML statechart of a Stakeholder, related to her/his participation to an assembly. This UML diagram is used with no modification. It represents the requirements that the proxies must be given before an assembly starts, that once a Stakeholder has registered to an assembly, s/he cannot delegate another, and that s/he can receive up to a maximum number of proxies. Note that, in this simplified model, delegations cannot be refused by the delegated shareholder, or withdrawn by the delegating shareholder.

The only Actors entitled to interact with the voting system are the Administrator and the Shareholders, hence the need of modifiers enforcing these constraints, that will be used by every public function, and that are shown in Table 3. Other modifiers are added at the bottom of Table 3, to account for other constraints common to more than one function.

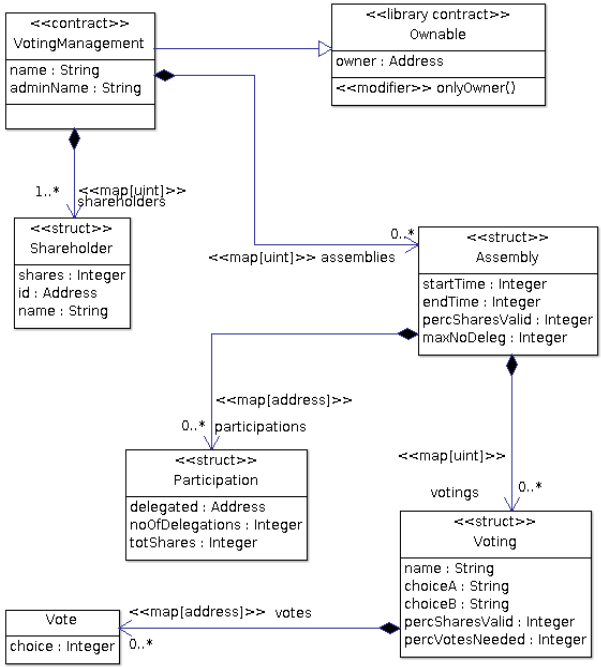

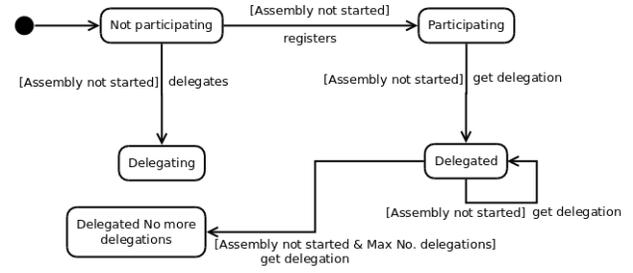

**Figure 5:** The modified UML diagram, showing the structure of the required SC of the voting system.

**Figure 6:** The statechart UML diagram, showing the state of a Stakeholder participating to an assembly, or delegating another Stakeholder.

**Table 3.** The modifiers of the SC.

| Modifier | Action - Notes |
|---|---|
| onlyOwner() | Enforces that the sender of the message is the owner of the contract (the Administrator). Inherited by Ownable standard contract. |
| onlyShareholder() | Enforces that the sender of the message is one of the shareholders registered in the contract. |
| OnlyOwnerOrShareholder() | Enforces that the sender of the message is the owner of the contract or one of the shareholders. |
| assemblyRunning() | Enforces that there is actually an assembly running at the time of the call. |
| AssemblyNotRunning() | Enforces that there is no assembly running at the time of the call. |

Finally, the public functions of the contract implement the functionalities described in the User Stories. They are summarized in Table 4. For each function we give the name (possibly followed by the kind of the function, in this case it can be "view") the modifiers enforcing the constraints related to its call, the parameters (which are Solidity types [13]), and a description of its purpose. In this example, the only function able to create a contract is the call to the constructor. In more elaborate cases – though not so common – a contract might create another contract.

The possible UML sequence diagrams showing the interactions among Actors and SCs are not representative in this example. In fact, all message calls happen between an Actor (Administrator or Stakeholder) and the SC. This simple example does not have direct interactions among more than two participants. For this reason, we do not report any sequence diagram.

**Table 4.** The public functions of the SC.

| Function | Modifiers, parameters | Action - Notes |
|---|---|---|
| constructor | string nameFirm string nameAdmin [(string nameSh, address addrSh, uint16 noShares)] | create the VotingManagement contract, inputting the name of the firm, the Administrator's name and, for each shareholder: name, address and number of shares. |
| addShareholder | onlyOwner string nameSh address addrSh uint16 noShares | Add a new shareholder, giving his name, address and number of shares. |
| Delete Shareholder | onlyOwner address addrSh | Delete the given shareholder, giving his address. Can be done only if the shareholder has no active participation in an assembly. |
| editShareholder | onlyOwner address addrSh string nameSh uint16 noShares | Update the given shareholder, giving his address (that cannot be changed), name and number of shares. Can be done only if the shareholder has no active participation in an assembly. |
| Change Administrator | onlyOwner address newOwner string nameAdmin | Give the address and the name of the new administrator. |
| Convene Assembly | onlyOwner | Convene an assembly, giving start and end date and time of the assembly, a short description, the minimum percentage of shares needed for its validity, and the maximum number of delegations that can be given to a single Shareholder. No existing assembly can overlap with the new one. |
| addVoting | onlyOwner | Add a call for voting to the given assembly, specifying the name of the voting, the two options that should be chosen, the minimum percentage of voting shares, and of votes needed to have a valid vote. The assembly must not have already started. |
| participate | onlyShareholder | Register the participation of the sender to the given Assembly, provided that the start date and time of the Assembly has not yet passed, and that the sender has not already delegated another Shareholder, or already registered. |

| Function | Modifiers, parameters | Action - Notes |
|---|---|---|
| delegate | onlyShareholder | Delegate his participation to a given Assembly to another Shareholder, provided that the start date and time of the Assembly has not yet passed, that the sender has not already registered his participation or delegated another Shareholder, that the delegated Shareholder has registered to the Assembly, and has not yet reached the maximum number of delegations. |
| castVote | onlyShareholder | Cast a vote for one of the choices of a given voting, provided that the sender is participating to the Assembly of the voting, that this Assembly has started and has not yet expired, and that the vote has not already cast. |
| verifyValidity *view* | OnlyOwnerOrShareholder | Read the total number of shares that participated to a given Assembly, and check if the minimum number has been reached. The Assembly must have expired. |
| readResults *view* | OnlyOwnerOrShareholder | Read the voting results (choice 1, choice 2 or no choice), given an Assembly, and the name of a voting. The Assembly must have expired. |
| deleteContract | onlyOwner | Permanently delete the contract. |

Starting from the steps shown before, it is easy to write the SC. Moreover, it does not entail Ether transfers, except for the GAS needed to execute the transactions, and its security issues are not relevant, owing to its simplicity and ease to check the preconditions of its messages.

For the sake of simplicity, we skip the design and implementation of the external system (step 6). It included the development of a responsive application holding the address and the private key of the Actors, and enabling them to send the proper messages to the SC running on the Ethereum Blockchain. This application has been developed using node.js, and web3.js on the client side. Web3.js is the Ethereum Javascript library used to communicate with the Ethereum Blockchain.

## 5 Conclusions

Despite the huge effort presently ongoing in developing DApps, software engineering practices are still poorly applied in software development of BOS. The field is in fact still in its infancy, and tools or techniques for modeling and managing the peculiarities a software developer must face when dealing with Blockchain Oriented software systems are still matter for researchers. Tools and techniques of traditional software engineering have not yet been adapted and modified to adhere to this new software paradigm. A sound software engineering approach might greatly help in overcoming many of the issues plaguing Blockchain development providing developers with instruments similar to those typically used in traditional software engineering to afford architectural design, security issues, testing planes and strategies and to improve software quality and maintenance. Researchers in software engineering have a big opportunity to start studying a field that is very important and brand new exploiting concepts, tools, instruments and ideas already consolidated in software engineering and changing and adapting them to this new software technology.

This work moves toward this direction providing a full modeling of interactions among traditional software and Blockchain environment, including Class diagrams, Statecharts, US's diagrams, Sequence diagrams, Smart Contracts diagrams, all for BOSE, as well as a general scheme for managing BOS development processes, and a practical example of a paradigmatic Blockchain Smart Contract implementing a voting system. Our work can be really valuable to Blockchain firms, including ICO startups, that could develop a competitive advantage using SE (BOSE) practices since the beginning.

## ACKNOWLEDGMENTS

This work was partially founded by the AIND project (Native Digital Administrations and Enterprises), funded by Sardinia Region, PIA call 2013, E.U. P.O. FESR 2007/2013, n. 3706 Rep. n. 316, 22/04/2016.